\documentclass[aps,prl,twocolumn,showpacs,10pt,superscriptaddress,preprintnumbers,nobibnotes]{revtex4-1}
\usepackage{amsmath}
\usepackage{graphicx}
\usepackage{amssymb}
\usepackage[colorlinks,citecolor=red]{hyperref}
\usepackage[thicklines]{cancel}
\usepackage{color}
\usepackage{ulem}
\usepackage{multirow}
\usepackage{booktabs}
\usepackage{ulem}
\usepackage{slashed}
\newcommand{\als}{\alpha_s}
\newcommand{\eps}{\epsilon}

\begin{document}
\title{Fully charm-tetraquark production at hadronic collisions with gluon radiation effects}
\author{Yefan Wang}
\email{wangyefan@nnu.edu.cn}
\affiliation{Department of Physics and Institute of Theoretical Physics, Nanjing Normal University, Nanjing,
Jiangsu 210023, China}
\affiliation{Nanjing Key Laboratory of Particle Physics and Astrophysics, Nanjing Normal University, Nanjing,
Jiangsu 210023, China}
\author{Ruilin Zhu}
\email{Corresponding author: rlzhu@njnu.edu.cn}
\affiliation{Department of Physics and Institute of Theoretical Physics, Nanjing Normal University, Nanjing,
Jiangsu 210023, China}
\affiliation{Nanjing Key Laboratory of Particle Physics and Astrophysics, Nanjing Normal University, Nanjing,
Jiangsu 210023, China}

\begin{abstract}
We report the first complete next-to-leading order  QCD calculation for processes involving fully charm tetraquark states, revealing that the renormalization constant of the color-singlet four charm quark operator is exactly unity at this order. We have investigated the possible quark configurations of the fully charm tetraquarks and expanded their states in the color symmetry-antisymmetry basis. By applying the transverse momentum dependent factorization formalism, large logarithms induced by soft and collinear gluon radiations are resummed to all orders in the expansion of the strong interaction coupling at the accuracy of next-to-leading logarithm.  By combining LHCb data on the total cross section of the exotic hadron $X(6900)$ and CMS measurements of its spin-parity, we extracted its nonperturbative but universal long-distance matrix element. The rapidity and transverse momentum distributions  of the $X(6900)$ and its spin-zero partners are also predicted, which await further experimental verification.
\end{abstract}
\maketitle
\textit{Introduction.}
Charm family hadrons, according to the color confinement principle, are believed to be capable of containing one, two, three, four or even more charm quarks. Although the $J/\psi$ meson composed of a charm-anticharm pair was discovered fifty-two years ago~\cite{E598:1974sol,SLAC-SP-017:1974ind} and the D meson  composed of a charm-antilight pair two years later~\cite{Goldhaber:1976xn}, it was not until 2020 that the LHCb experiment found the first evidence of a pure charm tetraquark state, the $X(6900)$, in the $J/\psi$-pair invariant mass spectrum using a data sample corresponding to an integrated luminosity of $9fb^{-1}$—thereby providing a signal for a hadron with multiple charm quarks~\cite{LHCb:2020bwg}.

Using a data sample focused on the small rapidity region,  both the ATLAS and CMS collaborations have confirmed the narrow $X(6900)$ state and observed another narrow structure around $7100MeV$ and a broad structure above twice the $J/\psi$ mass~\cite{ATLAS:2023bft,CMS:2023owd}.

Several particularly interesting aspects of these exotic states include: (1)the spin-parity quantum numbers for all observed data  are found to be compatible with the $J^{PC}=2^{++}$ hypothesis from the very recent CMS  analysis~\cite{CMS:2025fpt}, which constrains the possible internal quark configuration for these exotic structures; (2)a Regge trajectory is predicted and observed from their mass spectrum~\cite{Zhu:2020xni,Zhu:2024swp}, which gives hints of a radial or orbital excitation tetraquark family; (3)independent observation from three LHC experiments with different rapidity region reveals that the production rate of the fully charm tetraquark state is slightly affected by low and medium rapidity, yet its signal relative to that of $J/\psi$-pair is enhanced at medium and high transverse momentum~\cite{LHCb:2020bwg}.

Addressing and clarifying the above three points requires both a consistent theoretical framework and precise calculations. From the perspective of group theory, the hadron states can always be represented by a complete set of basis states. In terms of color space, one can choose either the color-symmetry-antisymmetry $\mathbf{6}\otimes \bar{\mathbf{6}}$/$\bar{\mathbf{3}}\otimes \mathbf{3}$ basis or the color-singlet-octet $\mathbf{1}\otimes \mathbf{1}$/$\mathbf{8}\otimes \mathbf{8}$ basis for local observations. In color-symmetry-antisymmetry basis, the two charm quarks can form the ``good''  axial-vector diquark with antisymmetric  color $\bar{3}$ and symmetric spin-flavor configuration or ``bad'' scalar diquark with symmetric  color $6$ and antisymmetric spin-flavor configuration. The two charmed antiquarks behave similar. Then charmed diquark and antidiquark carry color charge and form a color-charge-neutral and tightly-bound hadron from a QCD confining attractive force. Interestingly, charmed diquark here is distinctly different from light-diquark system, due to the additional constraint of isospin symmetry present in the latter.  In color-singlet-octet basis, the two  charm quark-antiquark pairs can form a color-singlet charmonium or a color-octet cluster separately, then the two charmonia or colored clusters form a color-charge-neutral and loosely-bound molecule hadron  from a residual strong interaction.

A true physical system may reside in one of the basis states mentioned above, or in a superposition of these basis states. For example, we can expand the fully heavy tetraquark state as
\begin{align}
|T_{4Q}\rangle =& \sum_{L=0}C^s_L
|(Q_1Q_2)_0;(\bar{Q}_3\bar{Q}_4)_0;L\rangle
\nonumber\\&+C^a_L
|(Q_1Q_2)_1;(\bar{Q}_3\bar{Q}_4)_1;L\rangle,
\end{align}
where $C^s_L$ and $C^a_L$ are the symmetric and antisymmetric configuration coefficients with $\sum_{L=0}|C^s_L|^2+|C^a_L|^2=1$, respectively. Orbital excitation effects
are marked with $L$. The higher Fock states with gluon content are omitted here but can be included in a similar manner.

In this Letter, we calculate the differential and total cross sections for the hadronic production of S-wave fully charm tetraquark states based on the color-symmetry-antisymmetry basis and nonrelativistic QCD (NRQCD) effective theory. Our calculations include contributions from all possible structures, such as $J^{PC}=0^{++}$ or $2^{++}$, as well as mixing effects arising from identical quantum numbers. At hadron colliders, gluon radiation effects are non-negligible; therefore, we provide a complete treatment of QCD corrections up to the next-to-leading order (NLO). Meanwhile, we perform transverse momentum resummation at the next-to-leading logarithmic (NLL) accuracy to mitigate large logarithmic terms induced by soft and collinear gluon emissions in the low transverse momentum $P_\perp$ region.

\textit{QCD resummation formulae.}
Using the QCD factorization and transverse momentum resummation theory~\cite{Collins:1984kg,CTEQ:1993hwr}, the differential cross section of fully charm tetraquark  at hadronic collisions can be factorized as
\begin{align}
&\frac{d\sigma(p+p\to T_{4c}+X)}{dydP_\perp^2}
\nonumber\\= &\int \frac{d^2 b\, e^{i\overrightarrow{P_\perp} \cdot \vec{b}}}{(2\pi)^2} \sum_{ij}\int^1_{x_A} \frac{d\xi_1}{\xi_1} \int^1_{x_B}  \frac{d\xi_2}{\xi_2}  f_i(\frac{x_A}{\xi_1}, \mu) f_j(\frac{x_B}{\xi_2}, \mu)\nonumber\\&\times
\frac{M^2}{S}\hat{\sigma}^{(0)}_{i'j'}\pi W^{NP}(b,M)C_{i'i}\left(\xi_1,b_*,\mu\right) C_{j'j}\left(\xi_2,b_*,\mu\right)\nonumber\\&\times e^{-\int_{C_1^2/b_*^2}^{C_2^2 M^2} \frac{d\bar{\mu}^2}{\bar{\mu}^2} \left[ A_{ij}(\alpha_s(\bar{\mu})) \ln \left( \frac{C_2^2 M^2}{\bar{\mu}^2} \right) + B_{ij}(\alpha_s(\bar{\mu})) \right]}\nonumber\\&+Y(P_\perp,M,x_A,x_B,\mu_F,\mu_R), \label{ResumFormulae}
\end{align}
where $x_{A(B)}=\frac{M}{\sqrt{S}}e^{\pm y}$ with the collider energy $\sqrt{S}$, the tetraquark mass $M$ and rapidity $y$. $f_i(x_i, \mu)$ is the parton distribution function. The matching scale is chosen as $\mu=C_3/b_*$ with $b_*=b/\sqrt{1+(b/b_{max})^2}$ such that there are no large logs in matching kernel $C_{ij}$~\cite{Catani:1989ne,Camarda:2021ict,Ebert:2020dfc,Balazs:1995nz,Monni:2016ktx,Catani:2014qha,Berger:1993yp,Zhu:2012ts,Bizon:2017rah,Ebert:2016gcn,Banfi:2014sua,Procura:2014cba,Catani:2013tia,Mueller:2013wwa,Becher:2012qa,deFlorian:2012mx,Kang:2017glf,Mantry:2009qz,Bozzi:2008bb,Banfi:2004yd,Landry:2002ix,Sun:2014dqm,Sun:2012vc,Chen:2018pzu,Kulesza:2002rh,Nadolsky:1999kb,Landry:1999an,Catani:1999hs,Laenen:1998qw,Balazs:1997hv,Ellis:1997ii,Kidonakis:1997gm,Contopanagos:1996nh,Cao:2018ntd,Bonvini:2012az,Beneke:2022obx,Boussarie:2023izj}. $C_i$ are on the order of ${\cal O}(1)$, and we have $C_1=C_3=2 e^{-\gamma_E}$, $C_2=1$. The  nonperturbative Sudakov factor $W^{NP}(b,M)=e^{-S_{NP}}$  with $S_{NP}=b^2\left(g_1 + g_2 \ln(M/(2Q_0)) + g_3 \ln(100x_1 x_2)\right)$ is employed with the known BLNY and SIYY form~\cite{Landry:2002ix,Sun:2014dqm,Sun:2012vc}, where $x_i=x_{A,B}/\xi_i$, \(g_1 = 0.181 \pm 0.005\),
\(g_2 = 0.167 \pm 0.01\), \(g_3 = 0.003\) with  \(b_{\text{max}} = 1.5 \, \text{GeV}^{-1}\), \(Q_0 = 1.55 \, \text{GeV}\) for quarks and  \(g_1 = 0.030 \pm 0.113\),
\(g_2 = 0.870 \pm 0.134\), \(g_3 = -0.170\pm0.030\)  with  \(b_{\text{max}} = 0.5 \, \text{GeV}^{-1}\), \(Q_0 = 1.6 \, \text{GeV}\) for gluon. The regular term is
\begin{align}
&Y(P_\perp,M,x_A,x_B,\mu_F,\mu_R) =\sum_{ij}\int^1_{x_A} \frac{d\xi_1}{\xi_1} \int^1_{x_B}  \frac{d\xi_2}{\xi_2} \nonumber\\ &\times  f_i(
\frac{x_A}{\xi_1}, \mu_F) f_j(\frac{x_B}{\xi_2}, \mu_F) R_{ij}(P_\perp,M,x_A,x_B,\mu_F,\mu_R).~
\end{align}

The above factorization formula is believed to hold to all orders in $\alpha_s$, and the short-distance perturbative coefficients can be calculated order by order in the perturbation QCD theory.  We expand these coefficients as

\begin{align}
A(\alpha_s(\mu)) &=\sum_n\left(\frac{\alpha_s(\mu)}{\pi}\right)^{(n)}A^{(n)},\nonumber\\ B(\alpha_s(\mu)) &=\sum_n\left(\frac{\alpha_s(\mu)}{\pi}\right)^{(n)}B^{(n)},\nonumber\\
C_{ij}\left(z,b,\mu\right) &=\sum_n\left(\frac{\alpha_s(\mu)}{\pi}\right)^{(n)}C_{ij}^{(n)},\nonumber\\
R_{ij}(P_\perp,M,x_A,x_B, \mu,\mu_R)&=\sum_n\left(\frac{\alpha_s(\mu)}{\pi}\right)^{(n)}R_{ij}^{(n)}.
\end{align}
Up to NLO+NLL accuracy, the nontrivial results are
\begin{align}
&A_{gg}^{(1)} =C_A,~~~~A_{q\bar{q}}^{(1)} =C_F,\nonumber\\
&B_{ij}^{(1)} =A_{ij}^{(1)}\ln\left(\frac{C_1^2e^{2\gamma_E}}{4C_2^2}\right)+\delta^{(1)}_{ij},~\text{where}~ ij=gg,q\bar{q},
\nonumber\\
&C_{gg}^{(0)}\left(z,b,\mu\right) =C_{q\bar{q}}^{(0)}\left(z,b,\mu\right) =\delta(1-z),\nonumber\\
&C_{i\bar{j}}^{(1)}\left(z,b,\mu\right) =P_{ij}(z)\ln\left(\frac{2e^{-\gamma_E}}{b\mu }\right) + \frac{A_{i\bar{j}}^{(1)}}{2}\delta(1 - z)
\nonumber\\&\quad\times\left[-\frac{\pi^2}{6} - 2\ln^2\left(\frac{C_1 e^{\gamma_E}}{2C_2}\right) + \ln\left(\frac{C_1 b\mu e^{2\gamma_E}}{4C_2}\right) \right.\nonumber\\&\quad\left.+Fin^{(1),(J)}_{i\bar{j},g_n}\left(\frac{C_1}{C_2 b},\mu\right)\right],~\text{where}~ i\bar{j}=gg,q\bar{q},\nonumber\\
&C_{qg}^{(1)}\left(z,b,\mu\right) =C_{\bar{q}g}^{(1)}\left(z,b,\mu\right)=P_{qg}(z)\ln\left(\frac{2e^{-\gamma_E}}{b\mu }\right),\nonumber\\&
C_{gq}^{(1)}\left(z,b,\mu\right) =C_{g\bar{q}}^{(1)}\left(z,b,\mu\right)=P_{gq}(z)\ln\left(\frac{2e^{-\gamma_E}}{b\mu }\right),\label{eq:hardkernel}
\end{align}
where $P_{ij}(z)$ is the parton splitting distribution function. $\delta^{(1)}_{gg}=\frac{11}{2}-\frac{2n_h+n_l}{3}$, $\delta^{(1)}_{q\bar{q}}=9-\frac{2n_h+2n_l}{3}$ with the heavy and light quark flavor numbers $n_h$ and $n_l$.
The expressions for $Fin^{(n),(J)}_{ij,g_n}\left(M,\mu\right) $ and
$R_{ij}^{(n)}$ are given in the Supplemental Material~\cite{sup.mat.}.
In above, $\hat{\sigma}^{(0)}_{ij} $ is the leading-order (LO) partonic cross section with
\begin{align}
\hat{\sigma}^{(0)}_{ij} = \frac{\pi}{M^4}
\frac{1}{c_i c_j s_i s_j}
\left| \mathcal{A}^{(0)}_{ij} \right|^2,
\end{align}
where the color and spin average factors are $c_g=N_c^2-1$, $c_q=c_{\bar{q}}=N_c$, $s_g=2-2\eps$, and $s_q=s_{\bar{q}}=2$. The Feynman amplitude squared from certain operator contribution can be written as
\begin{align}
\left| \mathcal{A}^{(n)}_{ij} \right|_{O^J_{g_1,g_2}}^2 = \frac{M\pi ^4 \als^4}{128 m_c^8} c^{(n)}_{ij,O^J_{g_1,g_2}}\langle0|O^{J}_{g_1,g_2}|0\rangle.
\end{align}
In the following, $m_c=M/4$ with the tetraquark mass $M=6847^{+44+48}_{-28-20}MeV$~\cite{CMS:2023owd} is adopted as the charm quark mass. $O^{J}_{g_1,g_2}$ are the NRQCD long-distance operators for
tetraquarks with spin quantum number $J$
\begin{align}
O_{g_1,g_2}^{J}=\mathcal{O}_{g_1}^{(J)} \sum_X\left|T_{4 c}^J+X\right\rangle\left\langle T_{4 c}^J+X\right| \mathcal{O}_{g_2}^{(J) \dagger}.
\end{align}
Therein $g_i$ can be either $\mathbf{6}\otimes \overline{\mathbf{6}}$ or $\overline{\mathbf{3}} \otimes \mathbf{3}$ in color symmetry-antisymmetry basis. Then we have~\cite{Bodwin:1994jh,Feng:2023agq}
\begin{align}
\mathcal{O}_{\overline{\mathbf{3}} \otimes \mathbf{3}}^{(0)} & =-\frac{1}{\sqrt{3}}\left[\psi_a^T\left(i \sigma^2\right) \sigma^i \psi_b\right]\left[\chi_c^{\dagger} \sigma^i\left(i \sigma^2\right) \chi_d^*\right] \mathcal{C}_{\overline{\mathbf{3}} \otimes \mathbf{3}}^{a b ; c d},\nonumber \\
\mathcal{O}_{\overline{\mathbf{3}} \otimes \mathbf{3}}^{(2;i j )} & =\left[\psi_a^T\left(i \sigma^2\right) \sigma^m \psi_b\right]\left[\chi_c^{\dagger} \sigma^n\left(i \sigma^2\right) \chi_d^*\right] \Gamma^{i j ; m n} \mathcal{C}_{\overline{\mathbf{3}} \otimes \mathbf{3}}^{a b ; c d},\nonumber \\
\mathcal{O}_{\mathbf{6} \otimes \overline{\mathbf{6}}}^{(0)} & =\left[\psi_a^T\left(i \sigma^2\right) \psi_b\right]\left[\chi_c^{\dagger}\left(i \sigma^2\right) \chi_d^*\right] \mathcal{C}_{\mathbf{6} \otimes \overline{\mathbf{6}}}^{a b ; c d},\label{NRQCDOperator}
\end{align}
where $\mathcal{O}_{\overline{\mathbf{3}} \otimes \mathbf{3}}^{(2)}=\frac{1}{2}\varepsilon^{*ij}\mathcal{O}_{\overline{\mathbf{3}} \otimes \mathbf{3}}^{(2;i j )} $ with the polarization tensor $\varepsilon^{ij}$, the rank-4 color tensor is defined as $\mathcal{C}_{\mathbf{6} \otimes \overline{\mathbf{6}}}^{cd;ef} = \frac{1}{2\sqrt{6}}\left(\delta_{ce}\delta_{df}+\delta_{cf}\delta_{de}\right)$ and
$\mathcal{C}_{\overline{\mathbf{3}} \otimes \mathbf{3}}^{cd;ef} = \frac{1}{2\sqrt{3}}\left(\delta_{ce}\delta_{df}-\delta_{cf}\delta_{de}\right)$. The rank-4 Lorentz tensor is defined as $\Gamma^{\alpha\beta ; \mu\nu}=\frac{1}{2}g^{\mu \alpha}g^{\nu \beta}+\frac{1}{2}g^{\mu \beta}g^{\nu \alpha}-\frac{1}{D-1}g^{\mu \nu}g^{\alpha\beta}$.

\begin{figure}[th]
\begin{center}
\includegraphics[width=0.12\textwidth]{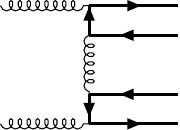}\,\,\,
\includegraphics[width=0.13\textwidth]{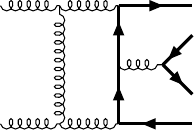}\,\,\,
\includegraphics[width=0.12\textwidth]{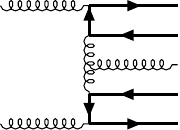}
\caption{Typical Feynman diagrams for the gluon-gluon fusion to fully charm tetraquark up to NLO. The thick black lines stand for the massive charm quarks. The hadronization process of the four charm quarks into a fully charm tetraquark is not displayed. The Feynman diagrams from the quark-antiquark annihilation or (anti)quark-gluon scattering can be plotted similarly.}
\label{fig:feynmandiagMI}
\end{center}
\end{figure}

The typical Feynman diagrams for fully charm tetraquark hadronic production are plotted in Fig.~\ref{fig:feynmandiagMI}. At LO, there are 62 and 4 tree-level non-vanishing Feynman diagrams contributing to $g+g\rightarrow T_{4c}$ and $q+\bar{q}\rightarrow T_{4c}$.
At NLO, there are 2008 and 170 one-loop non-vanishing Feynman diagrams contributing to $g+g\rightarrow T_{4c}$ and $q+\bar{q}\rightarrow T_{4c}$. Also we need to consider the real corrections and find  618, 98 and 98 tree-level non-vanishing Feynman diagrams contributing to $g+g\rightarrow T_{4c}+g$, $q+\bar{q}\rightarrow T_{4c}+g$ and $q+g\rightarrow T_{4c}+q$, respectively. Dimensional regularization with $D=4-2\eps$ is employed for both UV and IR divergences. We have generated and simplified the tree-level and one-loop amplitudes by using the mainstream packages~\cite{Hahn:2000kx,Shtabovenko:2020gxv,Shtabovenko:2023idz,Tkachov:1981wb,Chetyrkin:1981qh,Klappert:2020nbg,Shtabovenko:2016whf,Patel:2015tea}. In addition, the Ward identity is used to check our results.

At LO, the short-distance coefficients for gluon-gluon fusion are
\begin{align}
c^{(0)}_{gg,O^J_{g_1,g_2}}=\{\frac{15488}{27},2304,\frac{1408\sqrt{6}}{3}, \frac{200704}{27}\},
\end{align}
for $O^J_{g_1,g_2}=O^0_{\mathbf{6}\otimes \overline{\mathbf{6}},\mathbf{6}\otimes \overline{\mathbf{6}}},O^0_{\overline{\mathbf{3}} \otimes \mathbf{3},\overline{\mathbf{3}} \otimes \mathbf{3}},O^0_{\mathbf{6}\otimes \overline{\mathbf{6}},\overline{\mathbf{3}} \otimes \mathbf{3}},O^2_{\overline{\mathbf{3}} \otimes \mathbf{3},\overline{\mathbf{3}} \otimes \mathbf{3}}$, respectively. The result for $O^0_{\overline{\mathbf{3}} \otimes \mathbf{3},\mathbf{6}\otimes \overline{\mathbf{6}}}$ is the same as for $O^0_{\mathbf{6}\otimes \overline{\mathbf{6}},\overline{\mathbf{3}} \otimes \mathbf{3}}$. The non-vanishing contribution in quark-antiquark annihilation  for the spin-2 tetraquark is $c^{(0)}_{q\bar{q},O^2_{\overline{\mathbf{3}} \otimes \mathbf{3},\overline{\mathbf{3}} \otimes \mathbf{3}}}=32768/9$, while the result vanishes for the spin-0 tetraquark because of helicity conservation. We also find that the renormalization constants of all color singlet four charm quark operators in Eq.~(\ref{NRQCDOperator}) are exactly unity at NLO, which has never been explored in prior literature.

\begin{figure}[th]
\begin{center}
\includegraphics[width=0.35\textwidth]{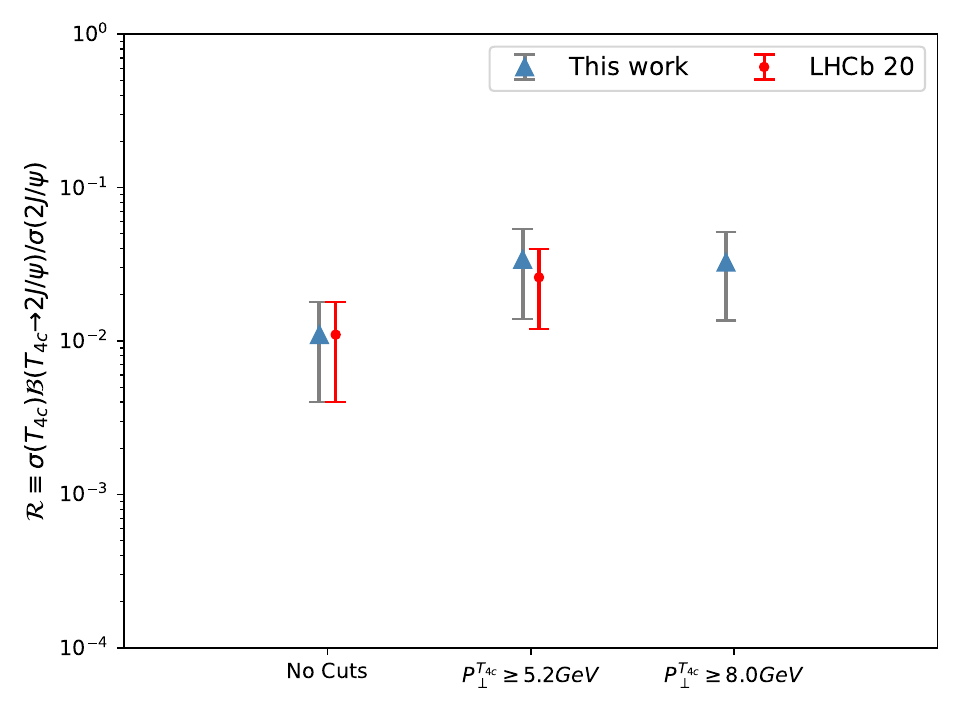}
\caption{ NLO prediction for cross-section of the $X(6900)$ relative to that of $J/\psi$-pair, times the branching fraction ${\cal B}(X(6900)\to J/\psi+J/\psi)$. The LHCb data is from Ref.~\cite{LHCb:2020bwg}.}\label{fig:Rratio}
\end{center}
\end{figure}

\textit{Constraint on long-distance matrix elements for fully charm tetraquark.}
The production cross-section of the $X(6900)$ relative to that of $J/\psi$-pair, times the branching fraction ${\cal B}(X(6900)\to J/\psi+J/\psi)$, was
measured to be ${\cal R}=[1.1\pm 0.4(stat)\pm0.3(syst)]\%$ without any $P^{T_{4c}}_{\perp}$ cuts in the proton-proton collision data at $\sqrt{S}=13TeV$ by the LHCb Collaboration~\cite{LHCb:2020bwg}. Subsequently, they have measured the production cross-section of the $J/\psi$-pair to be $\sigma(2J/\psi)=16.36\pm0.28(stat)\pm0.88(syst)nb$ at $\sqrt{S}=13TeV$ with both $J/\psi$ mesons in the transverse range $0<P_\perp^{J/\psi}<14GeV$ and rapidity range $2.0<y^{J/\psi}<4.5$~\cite{LHCb:2023ybt}. The related theoretical result for $\sigma(2J/\psi)$ appears in Refs.~\cite{He:2025kkw,Chapon:2020heu}. It is known that the spin-parity of the $X(6900)$ is determined as $2^{++}$ from the CMS data. From the above analysis, only one of  long-distance matrix elements (LDMEs) is required for a $J^{PC}=2^{++}$ fully charm tetraquark at leading power, which can be uniquely determined from the preceding experimental data. Combining the data and the NLO+NLL QCD formula, we determine the $J^{PC}=2^{++}$ LDMEs joint di-$J/\psi$ decay branching ratio as
\begin{align}
&\langle\mathcal{O}_{3\bar{3}}^{T4c}\rangle\left(^5S_2,2^{++}\right){\cal B}(T_{4c})=\langle0|O^2_{\overline{\mathbf{3}} \otimes \mathbf{3},\overline{\mathbf{3}} \otimes \mathbf{3}}|0\rangle {\cal B}(T_{4c})
\nonumber\\
&\quad\quad\quad=(2.22\pm0.80^{+1.29+0.62}_{-0.57-0.00})\times 10^{-4}GeV^9,
\end{align}
where the first column in uncertainties is from the existing LHCb measurements; the second column is from the scale uncertainty with $\mu_F=\mu_R \in \{M/2, M, 2M\}$ for $P_\perp$-integrated observables and $\{\sqrt{(M/2)^2+P_\perp^2},\allowbreak\sqrt{M^2+P_\perp^2},\allowbreak\sqrt{(2M)^2+P_\perp^2}\}$ for $P_\perp$-differential cross sections; and the  third column is from the parton distribution function sets: [PDF4LHC21\_mc~\cite{PDF4LHCWorkingGroup:2022cjn}, CT18NNLO~\cite{Hou:2019efy}, NNPDF23\_nnlo\_as\_0116~\cite{NNPDF:2014otw}] with
the LHAPDF package~\cite{Buckley:2014ana}.

Due to the lack of experimental data for $0^{++}$ fully charm tetraquark, other LDMEs in the paper are hard to determine reliably. As an order-of-magnitude estimate, using heavy quark spin symmetry, we can perform an estimation
\begin{align}
&\langle\mathcal{O}_{6\bar{6}}^{T4c}\rangle\left(^1S_0,0^{++}\right)\approx
\langle\mathcal{O}_{3\bar{3}}^{T4c}\rangle\left(^1S_0,0^{++}\right)\nonumber\\
\approx &
\langle\mathcal{O}_{mix}^{T4c}\rangle\left(^1S_0,0^{++}\right)\approx\frac{\langle\mathcal{O}_{3\bar{3}}^{T4c}\rangle\left(^5S_2,2^{++}\right)}{2J+1}\big|_{J=2}\nonumber\\
=&(4.44\pm1.60^{+2.58+1.24}_{-1.14-0.00})\times\frac{10^{-5}}{{\cal B}(T_{4c})}GeV^9.
\end{align}

To test the predictive power of the NLO+NLL QCD formula in other kinematic regions, we plot the ${\cal R}\equiv \sigma(T_{4c})/\sigma(2J/\psi) \mathcal{B}(T_{4c} \to 2J/\psi) $
with different transverse momentum cuts in Fig.~\ref{fig:Rratio}. From the plots, the ratio
${\cal R}$ increases when the lower limit of $P_\perp$ increases. Consider the $P^{2J/\psi}_\perp>5.2GeV$ case, the NLO+NLL calculation gives ${\cal R}^{Theo.}=[3.2\pm2.0]\%$, which shows good agreement with the LHCb data ${\cal R}^{LHCb}=[2.6\pm 0.6(stat)\pm0.8(syst)]\%$~\cite{LHCb:2020bwg}.
\begin{figure}[th]
\begin{center}
\includegraphics[width=0.35\textwidth]{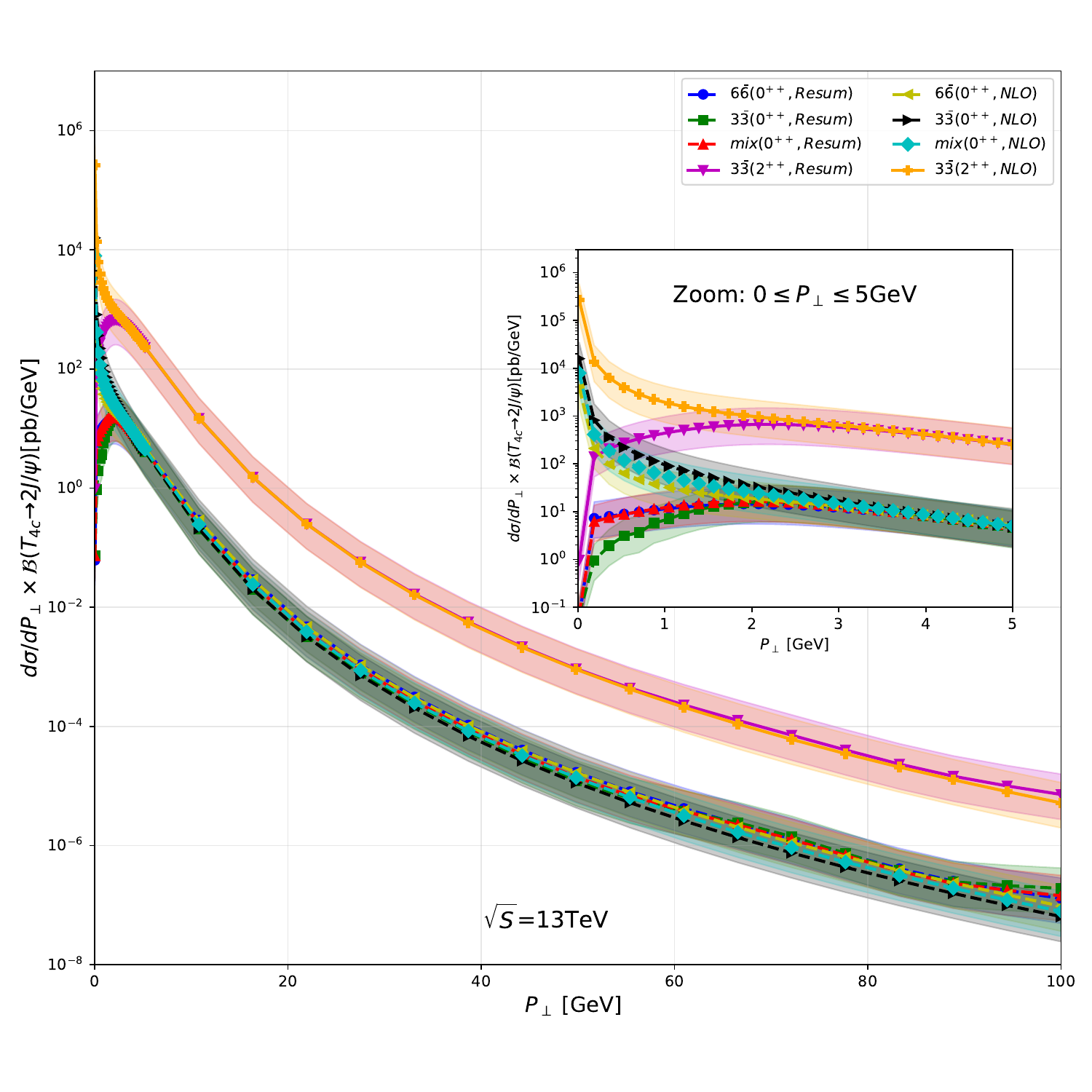}
\caption{ NLO and Resummation(NLO+NLL) differential cross-sections for the fully charm tetraquark with different quark configurations depending on the transverse momentum $P_\perp$, times the $2J/\psi$ decay branching fraction. In this paper $X(6900)$ is treated as a $2^{++}$ state and its $0^{++}$ partner is also plotted. The band is from the theoretical uncertainty.} \label{fig:transemom}
\end{center}
\end{figure}

\begin{figure}[th]
\begin{center}
\includegraphics[width=0.35\textwidth]{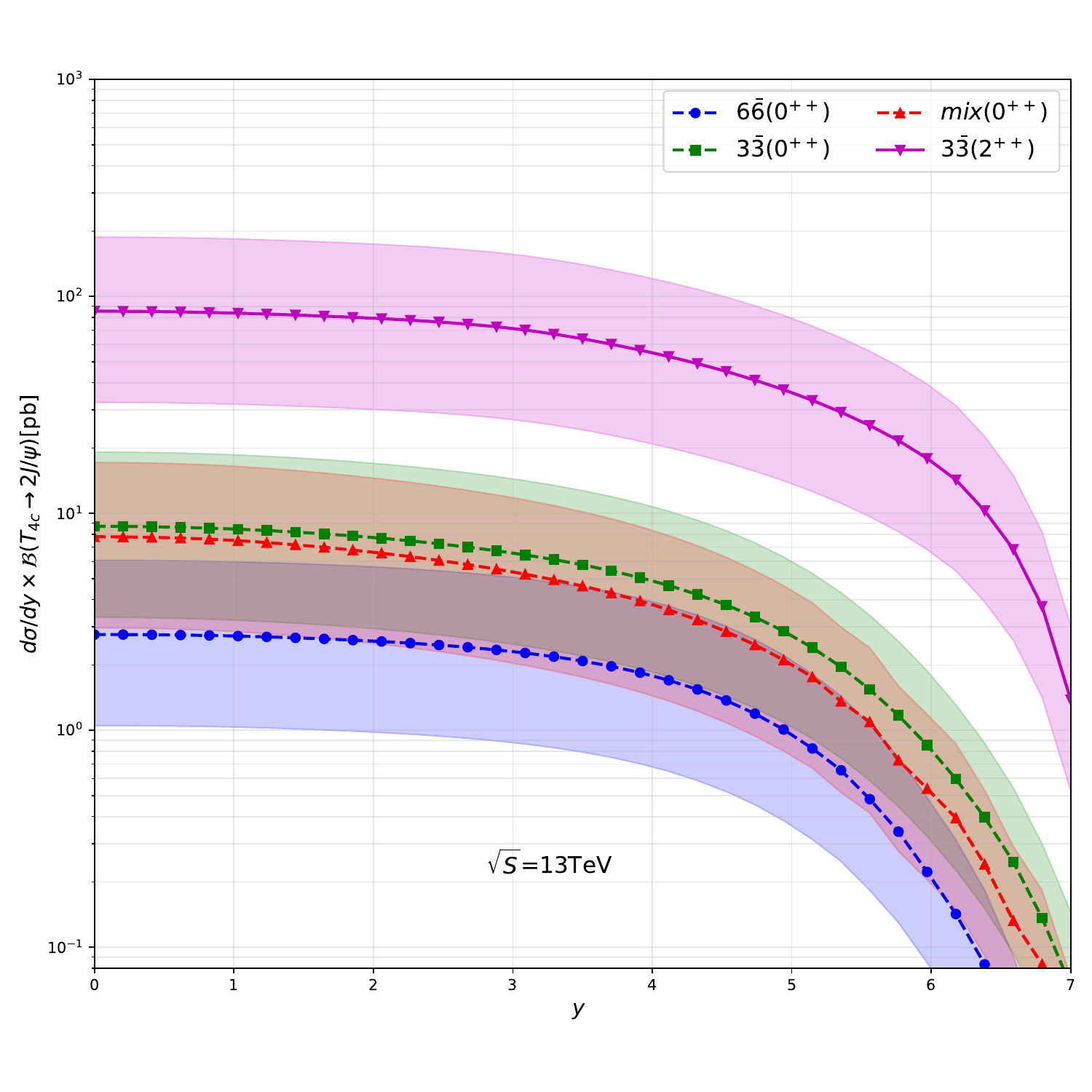}
\caption{ NLO differential cross-sections for the fully charm tetraquark with different quark configurations depending on the rapidity $y$, times the $2J/\psi$ decay branching fraction.}\label{fig:rapidity}
\end{center}
\end{figure}

\begin{figure}[th]
\begin{center}
\includegraphics[width=0.35\textwidth]{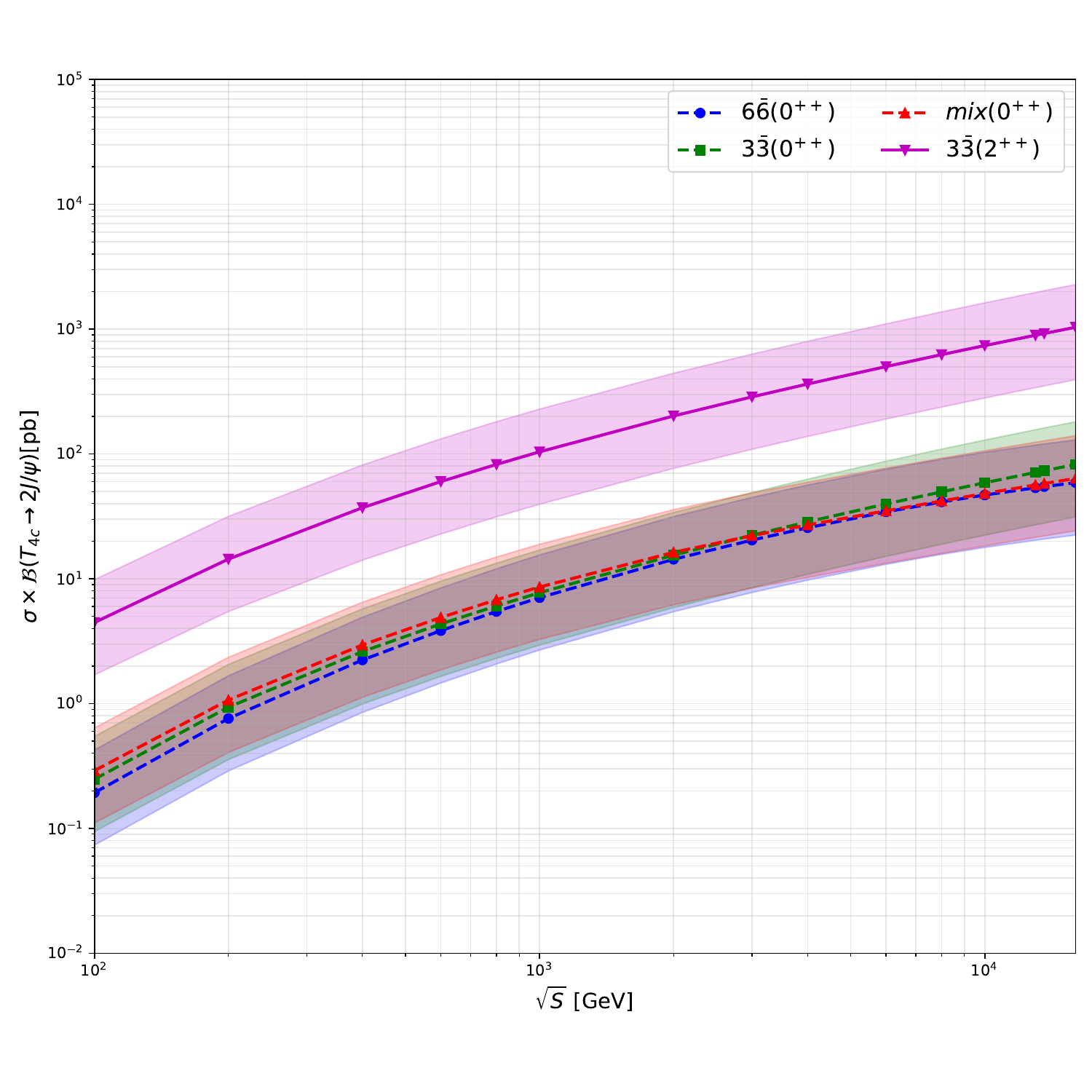}
\caption{NLO cross-section results for the fully charm tetraquark depending on the hadron-hadron center-of-mass energy $\sqrt{S}$, times the $2J/\psi$ decay branching fraction.  }\label{fig:crosssection}
\end{center}
\end{figure}

\textit{Differential and total production cross section.}
The soft and collinear gluon radiation induce singular structures specifically terms proportional to $1/P^2_\perp$ and $\ln(P^2_\perp/M^2)/P^2_\perp$ in the transverse momentum distribution, as shown in the divergence of NLO results with $P_\perp \ll M$ in Fig.~\ref{fig:transemom}. By applying the transverse momentum resummation formalism~\cite{Collins:1984kg,CTEQ:1993hwr,Catani:1989ne,Camarda:2021ict,Ebert:2020dfc,Balazs:1995nz,Monni:2016ktx,Catani:2014qha,Berger:1993yp,Zhu:2012ts,Bizon:2017rah,Ebert:2016gcn,Banfi:2014sua,Procura:2014cba,Catani:2013tia,Mueller:2013wwa,Becher:2012qa,deFlorian:2012mx,Kang:2017glf,Mantry:2009qz,Bozzi:2008bb,Banfi:2004yd,Landry:2002ix,Chen:2018pzu,Kulesza:2002rh,Nadolsky:1999kb,Landry:1999an,Catani:1999hs,Laenen:1998qw,Balazs:1997hv,Ellis:1997ii,Kidonakis:1997gm,Contopanagos:1996nh,Sun:2014dqm,Sun:2012vc,Cao:2018ntd,Bonvini:2012az,Beneke:2022obx,Boussarie:2023izj}, we improved the theoretical prediction of
fully charm tetraquark
on the kinematical distributions to sum over large logarithms $\ln(P^2_\perp/M^2)$ to all orders in the expansion of $\alpha_s$ at the NLO+NLL accuracy.

We also plot the rapidity distribution in Fig.~\ref{fig:rapidity} and the total cross section  in Fig.~\ref{fig:crosssection}  for different quark configurations. The differential cross section exhibits a relatively flat profile in rapidity, particularly across the low and intermediate rapidity ranges. From these plots, the $2^{++}$ tetraquark cross section times the $2J/\psi$ decay ratio is significantly enhanced  than that of the $0^{++}$ tetraquarks. This enhancement stems not only from differences in LDMEs, but also from the difference in short-distance coefficients, a finding which is also consistent with previous  theoretical and experimental works~\cite{CMS:2025fpt,Zhu:2020xni,Celiberto:2025ziy,Zhang:2020hoh,Feng:2020riv,Sang:2023ncm,Chen:2024orv,Belov:2024qyi}. Similar observation of enhancement of color antitriplet compared to other quark configurations is also shown in doubly charmed tetraquark system~\cite{Hyodo:2012pm}.

\begin{figure}[th]
\begin{center}
\includegraphics[width=0.35\textwidth]{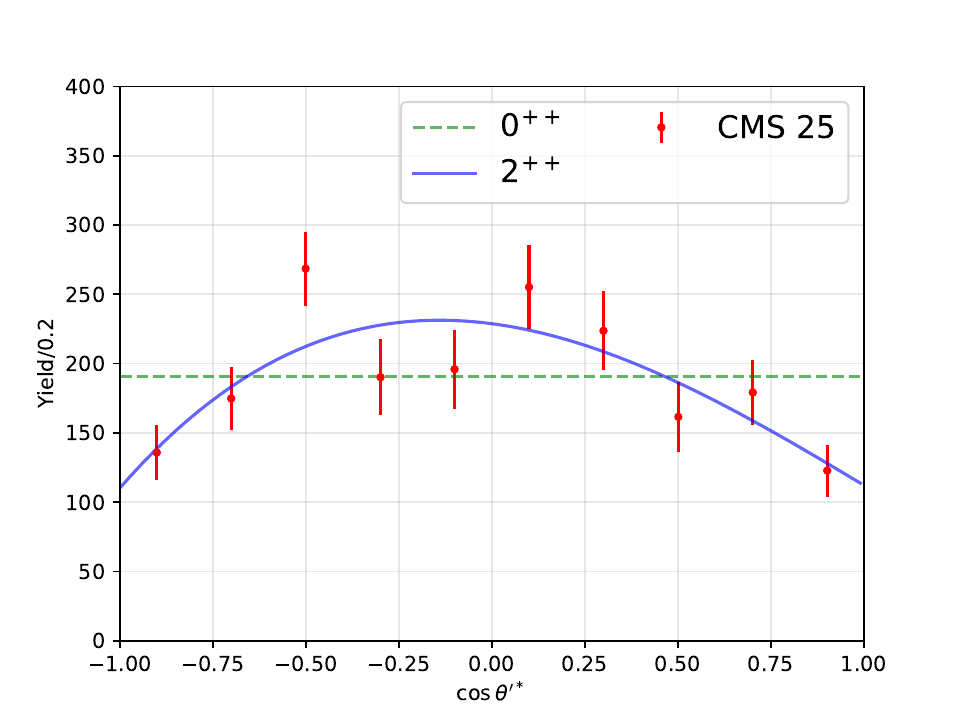}
\caption{ Distribution of the polar angle $\theta^{\prime *}$ between the $J/\psi$ and the fully charm tetraquark momenta. The data are from supplementary information  of Ref.~\cite{CMS:2025fpt}.} \label{fig:angle}
\end{center}
\end{figure}

Furthermore we consider the two-body decays of fully charm tetraquarks where the $0^{++}$ state will only produce an isotropic polar angular distribution, while the distribution for the $2^{++}$ state will be non-trivial. According to Jacob-Wick angular theory~\cite{Jacob:1959at}, we can always expand the polar angular distribution of the produced $2^{++}$ fully charm tetraquark state into $J/\psi$ pair as  $dN/d\cos\theta^{\prime *}=\Sigma_{i=1}^4 L_i\cos[(i-1)\theta^{\prime *}]$, where the coefficients $L_i$ depend on the production and decay mechanisms and $\theta^{\prime *}$ is the polar angle between one of $J/\psi$ and the fully charm tetraquark momenta. Fortunately, the CMS experiment has measured the polar angular distribution with $135fb^{-1}$, allowing us to fit these parameters. We show the results in Fig.~\ref{fig:angle}, the parameters for $2^{++}$ have been fitted as $ L_1=170.0(7.9)$, $ L_2=-8(13)$, $ L_3=-59(12)$, $ L_4=9(13)$ with $\chi^2/dof=1.8$, which determine the correlation between the production and decay of the fully charm tetraquark state. Due to the complexity of the CMS analysis where all the three peaks are included, we defer a detailed discussion to future work. The polar angle $\theta^{\prime *}$ distribution from the CMS experiment support the $J^{PC}=2^{++}$ assignment for the exotic structures in the $J/\psi$ pair final states.

\textit{Conclusion.}
We would like to conclude this work with the following remarks. First, our calculation on fully charm tetraquark is built upon the fundamental principles of QCD factorization and transverse momentum resummation, ensuring the rigor of the theoretical formalism and the thoroughness of the analysis. Second, we have for the first time constrained the non-perturbative but universal LDME of the
$X(6900)$ in a model-independent way by utilizing one LHCb data point, and based on this, the predicted ${\cal R}(P_\perp>5.2GeV)$-value agrees with the remaining experimental data.
This extraction can be also widely applied to other processes involving $X(6900)$ such as the production at electron-positron colliders. Third, for sufficiently large $P_\perp \gg M$, new factorization and resummation have been proposed in the literature~\cite{Kang:2014tta}, we leave this part for future work. Since the cross section is dominated in the small and moderate transverse momentum region, the impact on the extraction of LDMEs is negligible.  Fourth, although theoretical studies have been conducted on the spectrum of S-wave, P-wave, and higher excited fully charm tetraquark states~\cite{Wang:2023jqs,Wu:2024euj}, the work presented in this paper will provide an important foundation for systematically studying the production and inner structure of the family of fully charm tetraquark states. The reason is that we consider the symmetry-based basis expansion method to be effective, and after factorizing the short-distance coefficients and the long-distance matrix elements, the short-distance coefficients can be calculated order by order in perturbation theory. With the increasing data samples collected by the LHC, we also hope that experimental colleagues will perform more precise measurements of (differential) cross sections, rapidity and transverse momentum distributions for each exotic hadron in $J/\psi$-pair mass spectrum. Such efforts will provide further insights into clarifying the structural properties of these states, particularly regarding the existence of compact clustering mechanisms and the color confinement mechanism of four heavy quarks.

\textit{Acknowledgments.}
Thanks to Prof. Liu-Pan An, Prof. Zhi-Guo He and Mr. Kaiwen Chen for useful discussions. This work is supported by
the National Natural Science Foundation of China
Grants No.12322503, 12405117 and 12675120.

\begin{widetext}

\section*{Supplemental Material}

	In this supplemental material, we give the explicit analytic expressions for the matching kernel $Fin^{(1),(J)}_{i j,g_n}\left(M,\mu\right)$ and the regular function $R^{(1)}_{ij}$ in the main text. The     Feynman amplitude calculation is also detailed.
	
	\subsection{Matching kernel $Fin^{(1),(J)}_{i \bar{j},g_n}\left(M,\mu\right)$ }
	At QCD NLO+NLL accuracy, the nontrivial results for matching kernel $Fin^{(1),(J)}_{i \bar{j},g_n}\left(M,\mu\right)$ in Eq.~(5) of the main text are
	\begin{align}
		&Fin^{(1),(J=0)}_{gg,{\mathbf{6}} \otimes \overline{\mathbf{6}}}\left(M,\mu\right)=
		\left(\frac{11}{6}-\frac{2n_h+n_l}{9}\right)\ln\left(\frac{16\mu^2}{M^2}\right)
		+\frac{3689}{1584}\bigg(2 \text{Li}_2\left(-\frac{3}{5}\right)-2 \text{Li}_2\left(\frac{1}{5}\right)-\text{Li}_2\left(\frac{3}{5}\right)
		+\text{Li}_2\left(-\frac{1}{5}\right)
		\nonumber\\&\quad\quad\quad\quad\quad
		-\ln (3) \ln (5)\bigg)
		+\frac{10397}{2376} \text{Li}_2\left(-\frac{1}{3}\right)
		-\frac{9733}{4752}\text{Li}_2\left(\frac{1}{3}\right)
		-\frac{1975+3n_h}{792}\ln ^2\left(2 \sqrt{2}+3\right)+\frac{8-3n_h}{132}\ln ^2\left(7-4 \sqrt{3}\right)
		\nonumber\\&\quad\quad\quad\quad\quad
		+\frac{1229}{1056}\ln ^2(3)
		+\frac{2287+40n_h+16n_l}{2112} \pi ^2+\frac{4135-165n_h}{297 \sqrt{2}} \ln \left(2 \sqrt{2}+3\right)-\frac{10705-396n_l}{1782}\ln (2)
		\nonumber\\&\quad\quad\quad\quad\quad
		-\frac{493+458n_h+74n_l}{594}+{\cal O}(\eps).
		\\
		&Fin^{(1),(J=0)}_{gg,{\overline{\mathbf{3}}} \otimes \mathbf{3}}\left(M,\mu\right)=
		\left(\frac{11}{6}-\frac{2n_h+n_l}{9}\right)\ln\left(\frac{16\mu^2}{M^2}\right)
		+\frac{7}{1296}\bigg(2 \text{Li}_2\left(-\frac{3}{5}\right)-2 \text{Li}_2\left(\frac{1}{5}\right)-\text{Li}_2\left(\frac{3}{5}\right)+\text{Li}_2\left(-\frac{1}{5}\right)
		\nonumber\\&\quad\quad\quad\quad\quad
		-\ln (3) \ln (5)\bigg)
		+\frac{1613}{5832} \text{Li}_2\left(-\frac{1}{3}\right)
		-\frac{769}{11664}\text{Li}_2\left(\frac{1}{3}\right)
		-\frac{245+n_h}{648}\ln ^2\left(2 \sqrt{2}+3\right)-\frac{8+3n_h}{324}\ln ^2\left(7-4 \sqrt{3}\right)
		\nonumber\\&\quad\quad\quad\quad\quad
		+\frac{91}{864}\ln ^2(3)
		+\frac{2281+120n_h+48n_l}{15552} \pi ^2+\frac{343-117n_h}{729 \sqrt{2}} \ln \left(2 \sqrt{2}+3\right)+\frac{281+36n_l}{162}\ln (2)
		\nonumber\\&\quad\quad\quad\quad\quad
		+\frac{289-174n_h-78n_l}{486}+{\cal O}(\eps).
		\\
		&Fin^{(1),(J=0)}_{gg,mix}\left(M,\mu\right)=
		\left(\frac{11}{6}-\frac{2n_h+n_l}{9}\right)\ln\left(\frac{16\mu^2}{M^2}\right)
		+\frac{16639}{14256}\bigg(2 \text{Li}_2\left(-\frac{3}{5}\right)-2 \text{Li}_2\left(\frac{1}{5}\right)-\text{Li}_2\left(\frac{3}{5}\right)+\text{Li}_2\left(-\frac{1}{5}\right)
		\nonumber\\&\quad\quad\quad\quad\quad
		-\ln (3) \ln (5)\bigg)
		+\frac{149231}{64152} \text{Li}_2\left(-\frac{1}{3}\right)
		-\frac{135625}{128304}\text{Li}_2\left(\frac{1}{3}\right)
		-\frac{10235 + 19 n_h}{7128}\ln ^2\left(2 \sqrt{2}+3\right)
		\nonumber\\&\quad\quad\quad\quad\quad
		+\frac{64-57 n_h}{3564}\ln ^2\left(7-4 \sqrt{3}\right)+\frac{6031}{9504}\ln ^2(3)
		+\frac{105169+2280n_h+912n_l}{171072} \pi ^2+\frac{57709-2871n_h}{8019 \sqrt{2}} \ln \left(2 \sqrt{2}+3\right)
		\nonumber\\&\quad\quad\quad\quad\quad
		-\frac{423-44n_l}{198}\ln (2)
		-\frac{629+3018n_h+762n_l}{5346}+{\cal O}(\eps).
		\\
		&Fin^{(1),(J=2)}_{gg,\overline{\mathbf{3}} \otimes \mathbf{3}}\left(M,\mu\right)= \left(\frac{11}{6}-\frac{2n_h+n_l}{9}\right)\ln\left(\frac{16\mu^2}{M^2}\right)
		-\frac{31603}{6048}\text{Li}_2\left(\frac{1}{3}\right)-\frac{155}{1512}\text{Li}_2\left(-\frac{1}{3}\right)+\frac{(11+3n_h)}{112}\ln ^2\left(7-4 \sqrt{3}\right)
		\nonumber\\&\quad\quad\quad\quad\quad-\frac{\left(1109-19n_h\right)}{336}\ln ^2\left(2 \sqrt{2}+3\right)
		+\frac{\pi ^2 (-14251+720n_h)}{24192}
		-\frac{10741}{4032}\ln ^2(3)
		-\frac{11+3n_h}{6 \sqrt{3}}\ln \left(4 \sqrt{3}+7\right)
		\nonumber\\&\quad\quad\quad\quad\quad+\frac{(2027-6n_h)}{126 \sqrt{2}}\ln \left(2 \sqrt{2}+3\right)
		+\frac{(97655-3888n_l)}{13608} \ln (2)
		+\frac{611-1740n_h-912n_l}{4536}+{\cal O}(\eps).
	\end{align}
and
	\begin{align}
	&Fin^{(1),(J=2)}_{q\bar{q},{\overline{\mathbf{3}}} \otimes \mathbf{3}}\left(M,\mu\right)=
	\left(\frac{27}{4}-\frac{n_h+n_l}{2}\right)\ln\left(\frac{16\mu^2}{M^2}\right)
	+\frac{315}{128}\bigg(2 \text{Li}_2\left(-\frac{5}{3}\right)+2\text{Li}_2\left(\frac{1}{5}\right)+\text{Li}_2\left(\frac{3}{5}\right)
	\nonumber\\&\quad\quad\quad\quad\quad
	-\text{Li}_2\left(-\frac{1}{5}\right)-\text{Li}_2\left(\frac{1}{3}\right)+\ln^2(5)
	-\ln (3) \ln (5)\bigg)
	-\frac{263}{64} \text{Li}_2\left(-\frac{1}{3}\right)
	-\frac{211}{256}\ln^2(3)
	-\frac{675}{512} \pi ^2
	\nonumber\\&\quad\quad\quad\quad\quad
	-\frac{655-144n_l}{144}\ln (2)+\frac{91-16n_h-10n_l}{12}+{\cal O}(\eps).
	\end{align}
	In these formulae, the $n_l (n_h)$ are the numbers of the light (heavy) quarks in the quark loop. According to the QCD resummation formulae from Eq.~(2) of the main text,  the final result makes $M\to \frac{C_3}{b_*}=\frac{C_1}{C_2 b_*}$ and $\mu\to \frac{C_3}{b_*}$ after the CSS resummation procedure.
	
\subsection{Feynman amplitude calculation}	
The amplitude  of fully charm tetraquark production from the factorization theorem can be written as
\begin{align}
\mathcal{A} = \mathcal{T}_{ijkl}(c\bar{c}c\bar{c} )\langle T_{4c} +X |\overline{\psi}_i \psi_j \overline{\psi}_k \psi_l|0\rangle,
\end{align}
where $\mathcal{T}_{ijkl}(c\bar{c}c\bar{c})$ is the matrix element for the production of the four charm $c\bar{c}c\bar{c}$ final state with the charm-quark spinors removed, $i,j,k,l$ are spinor indices and $\psi$ is the charm fermion field operator. Inserting a complete set of four quark states, the amplitude becomes

\begin{align}
\mathcal{A} &= \sum_n \langle T_{4c}+ X | c\bar{c}c\bar{c} [n] \rangle \langle c\bar{c}c\bar{c} [n] |\overline{\psi}_i \psi_j \overline{\psi}_k \psi_l | 0 \rangle \mathcal{T}_{ijkl} \nonumber\\
&\equiv \sum_n \langle T_{4c} +X | c\bar{c}c\bar{c} [n] \rangle \text{Tr} \left[ \mathcal{T} \mathcal{P}^{(n)} \right].
\end{align}
Therein the projector \(\mathcal{P}^{(n)} \) projects four charm quarks onto the \(c\bar{c}c\bar{c} [n]\) state with certain  quantum numbers.  The production amplitude squared  of fully charm tetraquark can be rewritten as
\begin{align}
|\mathcal{A}|^2 &=   \sum_X\left|\sum_n \langle T_{4c} +X | c\bar{c}c\bar{c} [n] \rangle \text{Tr} \left[ \mathcal{T} \mathcal{P}^{(n)} \right]\right|^2\nonumber\\
&= \sum_X\left|\sum_n \langle T_{4c} +X | {\cal O}^{[n]~\dagger}_{4c}|0 \rangle\text{Tr} \left[ \mathcal{T} \mathcal{P}^{(n)} \right]\right|^2\nonumber\\
&= \sum_{m,m'} \sum_X\langle 0|{\cal O}^{[m]}|T_{4c} +X\rangle\langle T_{4c} +X | {\cal O}^{[m']~\dagger}_{4c}|0 \rangle\,\text{Tr} \left[ \mathcal{T} \mathcal{P}^{(m)} \right](\text{Tr} \left[ \mathcal{T} \mathcal{P}^{(m')} \right])^*\nonumber\\
&= \sum_{m,m'} \langle 0| O^{4c}_{m,m'}|0\rangle\,\text{Tr}\left[ \mathcal{T} \mathcal{P}^{(m)} \right](\text{Tr} \left[ \mathcal{T} \mathcal{P}^{(m')} \right])^*
\nonumber\\
&= \sum_{m,m'} \langle 0| O^{4c}_{m,m'}|0\rangle\,\mathcal{M}^m\left(\mathcal{M}^{m'}\right)^*,
\end{align}
where we have introduced the four charm quark operator \({\cal O}^{[n]~\dagger}_{4c}\) to create  the \(c\bar{c}c\bar{c} [n]\) state  from the vacuum, $\mathcal{M}$ represents the short-distance partonic amplitude which have been presented in the following $2\to 4$ and $2\to 5$ body processes. Unlike the two-body system of heavy quarkonium, in the four-quark system there is mixing between different operators already at leading power, thus necessitating a sum over indices $m$ and $m'$. In addition, the NRQCD long-distance operators for fully charm tetraquarks is introduced

\[
O^{4c}_{m,m'} = \sum_X  {\cal O}^{[m]}|T_{4c} +X\rangle\langle T_{4c} +X | {\cal O}^{[m']~\dagger}_{4c}.
\]

In paper we use the following notation instead of
 $O^{4c}_{n,n'}$ to create fully heavy tetraquarks with spin quantum number $J$ for convenience
\begin{align}
O_{g_1,g_2}^{J}=\mathcal{O}_{g_1}^{(J)} \sum_X\left|T_{4 c}^J+X\right\rangle\left\langle T_{4 c}^J+X\right| \mathcal{O}_{g_2}^{(J) \dagger},
\end{align}
where $g_i$ can be either $\mathbf{6}\otimes \overline{\mathbf{6}}$ or $\overline{\mathbf{3}} \otimes \mathbf{3}$ in color symmetry-antisymmetry basis. Then we have
\begin{align}
\mathcal{O}_{\overline{\mathbf{3}} \otimes \mathbf{3}}^{(0)} & =-\frac{1}{\sqrt{3}}\left[\psi_a^T\left(i \sigma^2\right) \sigma^i \psi_b\right]\left[\chi_c^{\dagger} \sigma^i\left(i \sigma^2\right) \chi_d^*\right] \mathcal{C}_{\overline{\mathbf{3}} \otimes \mathbf{3}}^{a b ; c d},\nonumber \\
\mathcal{O}_{\overline{\mathbf{3}} \otimes \mathbf{3}}^{(2;i j )} & =\left[\psi_a^T\left(i \sigma^2\right) \sigma^m \psi_b\right]\left[\chi_c^{\dagger} \sigma^n\left(i \sigma^2\right) \chi_d^*\right] \Gamma^{i j ; m n} \mathcal{C}_{\overline{\mathbf{3}} \otimes \mathbf{3}}^{a b ; c d},\nonumber \\
\mathcal{O}_{\mathbf{6} \otimes \overline{\mathbf{6}}}^{(0)} & =\left[\psi_a^T\left(i \sigma^2\right) \psi_b\right]\left[\chi_c^{\dagger}\left(i \sigma^2\right) \chi_d^*\right] \mathcal{C}_{\mathbf{6} \otimes \overline{\mathbf{6}}}^{a b ; c d},\label{NRQCDOperator}
\end{align}
where the rank-4 color tensor is defined as $\mathcal{C}_{\mathbf{6} \otimes \overline{\mathbf{6}}}^{cd;ef} = \frac{1}{2\sqrt{6}}\left(\delta_{ce}\delta_{df}+\delta_{cf}\delta_{de}\right)$ and
$\mathcal{C}_{\overline{\mathbf{3}} \otimes \mathbf{3}}^{cd;ef} = \frac{1}{2\sqrt{3}}\left(\delta_{ce}\delta_{df}-\delta_{cf}\delta_{de}\right)$. The rank-4 Lorentz tensor is defined as $\Gamma^{\alpha\beta ; \mu\nu}=\frac{1}{2}g^{\mu \alpha}g^{\nu \beta}+\frac{1}{2}g^{\mu \beta}g^{\nu \alpha}-\frac{1}{D-1}g^{\mu \nu}g^{\alpha\beta}$ with the space-time dimension $D$.

The partonic cross section becomes
\begin{align}
d\hat{\sigma}_{i+j\to T_{4c}[J,g_1,g_2]+X}
 =& \frac{1}{2\hat{s}}\left( \prod_{i=1}^{n} \frac{d^{D-1}p_i}{(2\pi)^{D-1} 2E_i} \right) (2\pi)^D \delta^{(D)}\!\left( k_1 + k_2 - \sum_{i=1}^{n} p_i \right) \left| \mathcal{A} \right|^2\nonumber\\ = &
  d\overline{\hat{\sigma}}(c\bar{c}c\bar{c}[J,g_1,g_2] + X;\mu_R,\mu'_F) \langle 0|O_{g_1,g_2}^{J}|0 \rangle(\mu'_F).
\end{align}
Therein the partonic differential cross section $d\hat{\sigma}_{i+j\to T_{4c}+X}(\mu_R, \mu_F, x_1, x_2, \hat{s})$ can be factored into long-distance matrix elements (LDMEs) $\langle 0|O_{g_1,g_2}^{J}|0 \rangle(\mu'_F)$ and short-distance matrix coefficients (SDCs) $d\overline{\hat{\sigma}}(c\bar{c}c\bar{c}[J,g_1,g_2] + X;\mu_R,\mu'_F) $. The latter can be calculated perturbatively in terms of the strong coupling constant $\als$.  The renormalization scale $\mu_R$ dependence is detailed in the above. In general, both the
LDMEs and SDCs are dependent on the new factorization scale $\mu'_F$ in NRQCD effective theory. However, this paper shows that this new factorization scale $\mu'_F$ dependence is absent at Next-to-leading order(NLO).
However, we can expect that this new factorization scale dependence will appear at Next-to-next-to-leading order(NNLO), just like the anomalous dimension of the $J/\psi$ and $B_c$ decay constant appears at NNLO.

For the SDCs calculation, there are two processes in the leading order (LO). At the quark level, we have
\begin{align}
g(k_1)+g(k_2) \rightarrow c(p_1)+c(p_2)+\bar{c}(p_3)+\bar{c}(p_4).
\end{align}
and
\begin{align}
q(k_1)+\bar{q}(k_2) \rightarrow c(p_1)+c(p_2)+\bar{c}(p_3)+\bar{c}(p_4).
\end{align}
In the NLO, in addition to the one-loop virtual corrections to the above processes, the following $2\to 5$ body processes are included as real corrections. At the quark level, we have
\begin{align}
g(k_1)+g(k_2) &\rightarrow c(p_1)+c(p_2)+\bar{c}(p_3)+\bar{c}(p_4)+g(p_5), \nonumber\\
q(k_1)+\bar{q}(k_2) &\rightarrow c(p_1)+c(p_2)+\bar{c}(p_3)+\bar{c}(p_4)+g(p_5), \nonumber\\
q(k_1)+g(k_2) &\rightarrow c(p_1)+c(p_2)+\bar{c}(p_3)+\bar{c}(p_4)+q(p_5), \nonumber\\
\bar{q}(k_1)+g(k_2) &\rightarrow c(p_1)+c(p_2)+\bar{c}(p_3)+\bar{c}(p_4)+\bar{q}(p_5).
\end{align}

We assume that the fully charm state $T_{4c}$ are constructed by diquark-antidiquark basis. Then the diquark can be spin-singlet or spin-triplet. For the spin-singlet diquark contribution, the state of $c(p_1)c(p_2)\bar{c}(p_3)\bar{c}(p_4)$ are enforced in $J^{PC} = 0^{++}$ by the replacements
\begin{align}
&\bar{u}(p_1)X_1v(p_3)\bar{u}(p_2)X_2v(p_4)\rightarrow -\text{Tr}\left[\Pi_{0} X_2\Pi_{0} X_1^C\right]\mathcal{C}_{6\otimes\bar{6}}^{cd;ef},\nonumber\\
&\bar{u}(p_1)X_1v(p_4)\bar{u}(p_2)X_2v(p_3)\rightarrow \text{Tr}\left[\Pi_{0} X_2\Pi^{C}_{0} X_1^C\right]\mathcal{C}_{6\otimes\bar{6}}^{cd;ef},
\label{eq:rep0}
\end{align}
where the superscript $C$ stands for the charge conjugate, and the spin-singlet projector is
\begin{align}
\Pi_{0}= \frac{\gamma_{5}\left(\frac{\slashed{P}}{4}+m_c\right)}{\sqrt{2}},
\end{align}
where $P$ is the momentum of the $T_{4c}$. And the color projection tensors is
\begin{align}
\mathcal{C}_{\mathbf{6} \otimes \overline{\mathbf{6}}}^{cd;ef} &= \frac{1}{2\sqrt{6}}\left(\delta_{ce}\delta_{df}+\delta_{cf}\delta_{de}\right).
\end{align}
In the lowest order of velocity, the relative velocity between quarks is zero, then we have
\begin{align}
p_1=p_2=p_3=p_4=\frac{P}{4}.
\end{align}
and
\begin{align}
P^2= M^2 = (4m_c)^2.
\end{align}

While the spin-triplet diquark can construct the $T_{4c}$ with $J^{PC}$ = $0^{++}$ or $2^{++}$. These contributions can be obtained by the replacements
\begin{align}
&\bar{u}(p_1)X_1v(p_3)\bar{u}(p_2)X_2v(p_4)\rightarrow -\text{Tr}\left[\Pi_{1\mu} X_2\Pi_{1\nu} X_1^C\right]\mathcal{C}_{\bar{3}\otimes3}^{cd;ef}J^{\mu \nu}_{0,2},\nonumber\\
&\bar{u}(p_1)X_1v(p_4)\bar{u}(p_2)X_2v(p_3)\rightarrow \text{Tr}\left[\Pi_{1\mu} X_2\Pi^{C}_{1\nu} X_1^C\right]\mathcal{C}_{\bar{3}\otimes3}^{cd;ef}J^{\mu \nu}_{0,2},
\label{eq:rep2}
\end{align}
where the spin-triplet projector is
\begin{align}
\Pi_{1,\mu}= \frac{\gamma_{\mu}\left(\frac{\slashed{P}}{4}+m_c\right)}{\sqrt{2}}.
\end{align}
The color projection tensors is
\begin{align}
\mathcal{C}_{\overline{\mathbf{3}} \otimes \mathbf{3}}^{cd;ef} &= \frac{1}{2\sqrt{3}}\left(\delta_{ce}\delta_{df}-\delta_{cf}\delta_{de}\right).
\end{align}
And the covariant projectors $J^{\mu \nu}_{0,2}$ in $D$-dimension are
\begin{align}
J^{\mu \nu}_{0}&=\frac{1}{\sqrt{D-1}}\eta^{\mu \nu},\nonumber\\
J^{\mu \nu}_{2}&=\epsilon_{\alpha\beta}\left[\frac{1}{2}\eta^{\mu \alpha}\eta^{\nu \beta}+\frac{1}{2}\eta^{\mu \beta}\eta^{\nu \alpha}-\frac{1}{D-1}\eta^{\mu \nu}\eta^{\alpha\beta}\right],
\end{align}
where $\epsilon_{\alpha\beta}$ is the polarization tensor and
\begin{align}
\eta^{\mu \nu} = -g^{\mu\nu} + \frac{P^\mu P^\nu}{M^2}.
\end{align}

\subsection{Regular term $Y$}
	
The four-momentum of the fully charm tetraquark  in the hadronic frame (Center-of-mass frame of hadrons with  $K_1+K_2=(\sqrt{S},\vec{0},0$) ) can be expressed in terms of its transverse momentum \( P_\perp \) and the rapidity \( y \) as,
\begin{align}
P^\mu = (M_T \cosh y, \vec{P}_\perp, M_T \sinh y), \label{eq:etaq_momentum}
\end{align}
where \( M_T = \sqrt{M^2 + P_\perp^2} \) with the on-shell condition \( P^2 = M^2 \).

Similarly, in the partonic frame (Center-of-mass frame of partons with $k_1+k_2=((x_1+x_2)\frac{\sqrt{S}}{2},\vec{0},(x_1-x_2)\frac{\sqrt{S}}{2})$), the  four-momentum of the fully charm tetraquark can be written as
\begin{align}
P'^\mu = (M_T \cosh y', \vec{P}_\perp, M_T \sinh y'). \label{eq:etaq_momentum}
\end{align}

These two rapidities are related by
\begin{align}
y = y' + \tilde{y}=y'+\frac{1}{2}\ln\left(\frac{x_1}{x_2}\right), \label{eq:rapidity_relation}
\end{align}
where \( \tilde{y} \) is the Lorentz boost term  entirely due to the ratio of \( x_1 \) and \( x_2 \).

At LO, the fully charm tetraquark is fixed at $P_\perp=0$ and $y'=0$ from the momentum conservation with $P=k_1+k_2$  in the center-of-mass frame of two incoming partons. Thus the rapidity of fully charm tetraquark in hadron frame
is $y=\tilde{y}=\frac{1}{2}\ln\left(\frac{x_1}{x_2}\right)$.

The differential cross section at LO can be written as

\begin{align}
\frac{d\sigma(p+p\to T_{4c}+X)}{dydP_\perp^2}|^{LO}= & \sum_{ij}\int^1_{x_A} \frac{d\xi_1}{\xi_1} \int^1_{x_B}  \frac{d\xi_2}{\xi_2}  f_i(\frac{x_A}{\xi_1}, \mu_F) f_j(\frac{x_B}{\xi_2}, \mu_F)  \frac{M^2}{S} \hat{\sigma}^{(0)}_{ij}\delta(1-\xi_1)\delta(1-\xi_2)\delta(P_\perp^2)\nonumber\\=&\sum_{ij}\frac{M^2}{S}\hat{\sigma}^{(0)}_{ij} f_i(x_A, \mu_F) f_j(x_B, \mu_F) \delta(P_\perp^2),
\end{align}
where
\begin{align}
&\xi_1=\frac{x_A}{x_1},~ \xi_2=\frac{x_B}{x_2},~ x_A=\frac{M}{\sqrt{S}}e^y,~ x_B=\frac{M}{\sqrt{S}}e^{-y}.
\end{align}

At NLO, the fully charm tetraquark will have a nontrivial transverse momentum $P_\perp$ and rapidity $y'$ in partonic frame because a new particle is emitted.
The differential cross section from the NLO real corrections (NLO,R) can be written as
\begin{align}
\frac{d\sigma(p+p\to T_{4c}+X)}{dydP_\perp^2}|^{NLO,R} = \sum_{ij}\int dx_1 \int dx_2 f_i(x_1, \mu_F) f_j(x_2, \mu_F) \hat{s} \frac{d\hat{\sigma}_{2\to2}}{d\hat{t}} \delta(\hat{s} + \hat{t} + \hat{u} - M^2),
\end{align}
where
\begin{align}
\frac{d\hat{\sigma}_{2\to2}}{d\hat{t}} = \frac{1}{16\pi\hat{s}^2}\frac{1}{\Gamma(1-\epsilon)}\left(\frac{4\pi\mu_R^2}{P_\perp^2}\right)^{\epsilon}|\overline{{\cal A}_{2\to2}}|^2.
\end{align}
If we let the transverse momentum approach zero, i.e. \(P_\perp\to 0\), the differential cross section will diverge. The reason is due to the radiation of soft and collinear partons, which will produce infrared divergences. Therefore, we can first extract the singular terms of the differential cross section.

We rewrite the parameters as follows
\begin{align}
&\hat{s}=\frac{M^2}{\xi_1 \xi_2},~ \hat{t}=M^2\left(1-\frac{x_T}{\xi_1}\right),~ \hat{u}=M^2\left(1-\frac{x_T}{\xi_2}\right),~ x_T=\sqrt{1+\frac{P_\perp^2}{M^2}}.
\end{align}
In the soft and collinear limit, the transverse momentum approaching to zero, i.e. \(P_\perp\to 0\), using the above expression we have
\begin{align}
\delta(\hat{s} + \hat{t} + \hat{u} - M^2) &=\frac{\xi_1 \xi_2}{M^2 x_T^2}\delta\left(\left(1-\frac{\xi_1}{x_T}\right)\left(1-\frac{\xi_2}{x_T}\right)-\frac{P_\perp^2/M^2}{1+P_\perp^2/M^2}\right)\nonumber\\
&\xrightarrow{P_\perp\to 0} \frac{\xi_1 \xi_2}{M^2}\delta\left(\left(1-\xi_1\right)\left(1-\xi_2\right)-\frac{P_\perp^2}{M^2}\right)\nonumber\\
&\xrightarrow{P_\perp\to 0} \frac{\xi_1 \xi_2}{M^2}\left[\frac{\delta(1-\xi_1)}{(1-\xi_2)_+}+\frac{\delta(1-\xi_2)}{(1-\xi_1)_+}-\ln\left(\frac{P_\perp^2}{M^2}\right)\delta(1-\xi_1)\delta(1-\xi_2)\right].
\end{align}
In the above limit case, the  differential cross section from the NLO real corrections becomes
\begin{align}
\frac{d\sigma(p+p\to T_{4c}+X)}{dydP_\perp^2}|^{NLO,R}_{Singular}&=
\left(\frac{d\sigma}{dydP_\perp^2}|^{NLO,R}\right)_{P_\perp\to 0}\nonumber\\ &= \sum_{ij}\int \frac{dx_1}{x_1} \int \frac{dx_2}{x_2}  f_{i'}(x_1, \mu_F) f_{j'}(x_2, \mu_F)  \frac{M^2}{S} \hat{\sigma}^{(0)}_{ij}\frac{\alpha_s(\mu_F)}{2\pi}\frac{1}{\Gamma(1-\epsilon)}\left(\frac{4\pi\mu_R^2}{P_\perp^2}\right)^{\epsilon}\nonumber\\&\times\frac{1}{P_\perp^2}\left[P_{jj'}(\xi_2)\delta_{ii'}\delta(1-\xi_1)+P_{i'i}(\xi_1)\delta_{j'j}\delta(1-\xi_2)\right.\nonumber\\&\left.-2C_{i'j'}\delta_{i'i}\delta_{j'j}\ln\left(\frac{P_\perp^2}{M^2}\right)\delta(1-\xi_1)\delta(1-\xi_2)\right].
\end{align}
From above, the expression will diverge when
\(P_\perp\to 0\). Note that the splitting distribution functions in D-dimension is confirmed in the derivation.  In addition, one can easily change the integration variable via substitution as
$\int \frac{dx_1}{x_1} \int \frac{dx_2}{x_2} =\int \frac{d\xi_1}{\xi_1} \int \frac{d\xi_2}{\xi_2}$.

We can perform the Fourier transformation on the differential cross section into the impact parameter $b$-space
\begin{equation}
\begin{aligned}[b]
\left. W(b) \right|_{\mathrm{Singular}}^{NLO,R}
={}& \int d^2 P_\perp\, e^{-i\vec P_\perp\cdot\vec b}
\left.\frac{d\sigma(p+p\to T_{4c}+X)}{dy\,d^2P_\perp}\right|_{\mathrm{Singular}}^{NLO,R}
\\[2pt]
={}& \sum_{ij}\int_{x_A}^{1}\frac{d\xi_1}{\xi_1}
\int_{x_B}^{1}\frac{d\xi_2}{\xi_2}
 f_{i'}\!\left(\frac{x_A}{\xi_1},\mu_F\right)
 f_{j'}\!\left(\frac{x_B}{\xi_2},\mu_F\right)
 \frac{M^2}{S}\,\hat\sigma_{ij}^{(0)}
 \frac{\alpha_s(\mu_F)}{2\pi}\,\Gamma(1+\epsilon)(4\pi)^\epsilon\pi
\\[2pt]
&\times\Bigg[
 P_{jj'}(\xi_2)\,\delta_{i'i}\,\delta(1-\xi_1)
 \left(-\frac{1}{\epsilon}+\ln\!\left(\frac{4e^{-2\gamma_E}}{b^2\mu_R^2}\right)\right)
 + (\xi_1\leftrightarrow\xi_2,\ i\leftrightarrow j,\ i'\leftrightarrow j')
\\[2pt]
&\qquad +2C_{i'j'}\,\delta_{i'i}\,\delta_{j'j}\,
 \delta(1-\xi_1)\,\delta(1-\xi_2)
 \Bigg(
 \frac{1}{\epsilon^2}
 +\frac{1}{\epsilon}\ln\!\left(\frac{\mu_R^2}{M^2}\right)
 +\frac{1}{2}\ln^2\!\left(\frac{\mu_R^2}{M^2}\right)
\\[2pt]
&\qquad\quad
 -\frac{\pi^2}{6}
 -\frac{1}{2}\ln^2\!\left(\frac{M^2b^2e^{2\gamma_E}}{4}\right)
 \Bigg)\Bigg].
\end{aligned}
\end{equation}

The difference between the complete NLO real corrections and the NLO real singular part (The asymptotic results of the complete NLO real corrections when $P_\perp\to 0$) is
\begin{equation}
Y(P_\perp,M,x_A,x_B,\mu_F,\mu_R)
=
\left.\frac{d\sigma(p+p\to T_{4c}+X)}{dy\,dP_\perp^2}\right|^{NLO,R}
-
\left.\frac{d\sigma(p+p\to T_{4c}+X)}{dy\,dP_\perp^2}\right|_{\mathrm{Singular}}^{NLO,R}.
\end{equation}

Because there is no divergence in the regular part of the NLO real corrections, this term does not require any processing. This is also the regular $Y$-term in Eq.~(2) of the main text.

We found that 618, 98 and 98 tree-level non-vanishing Feynman diagrams contributing to $g+g\to T_{4c}+g$, $q+\bar q\to T_{4c}+g$ and $q+g\to T_{4c}+q$ (also for $\bar q+g\to T_{4c}+\bar q$), respectively. We give the analytic expression of $\left|\mathcal{M}^{(1)}_{ijk}\right|^2(z,y_\theta)$ in the attached Mathematica notebook files since they are too lengthy.

Then the regular function $R_{ij}$ in the main text where the renormalization and factorization scales are not explicitly shown can be obtained as follows
\begin{equation}
R_{ij}(P_\perp,M,x_A,x_B,\mu_F,\mu_R)
= \sum_{n}\left(\frac{\alpha_s(\mu_F)}{\pi}\right)^{(n)}	R_{ij}^{(n)}(P_\perp,M,x_A,x_B,\mu_F,\mu_R).
\end{equation}
And we have the nontrivial result at NLO
\begin{equation}
\begin{aligned}[b]
&R_{ij}^{(1)}(P_\perp,M,x_A,x_B,\mu_F,\mu_R)
={}\frac{1}{2\pi}\Bigg(
\frac{1}{8\alpha_s M^2\hat\sigma_{i'j'}^{(0)}\,c_i c_j s_i s_j}
\left|\mathcal{M}^{(1)}_{ijk}\right|^2(z,y_\theta)
\delta(\hat s+\hat t+\hat u-M^2)
\\
&\,
-\frac{1}{P_\perp^2}\Bigg\{
P_{i'i}(\xi_1)\delta(1-\xi_2)
+\delta(1-\xi_1)P_{j'j}(\xi_2)
+2C_{ij}\delta(1-\xi_1)\delta(1-\xi_2)
\ln\!\left(\frac{M^2}{P_\perp^2}\right)
\Bigg\}\Bigg).
\end{aligned}
\end{equation}
where the nonvanishing $C_{ij}$ are $C_{gg}=C_A$ and $C_{q\bar q}=C_F$. In addition $y_\theta=\cos\theta$ with the angle $\theta$ between the momenta of initial parton $i$ and the tetraquark. The color and spin average factors are $c_g=N_c^2-1$, $c_q=c_{\bar q}=N_c$, $s_g=2-2\epsilon$, and $s_q=s_{\bar q}=2$. The parameter $z=M^2/\hat s$ with the partonic center-of-mass energy $\hat s$. The remaining Mandelstam variables are related by $\hat t=-\frac{\hat s}{2}(1-z)(1-y_\theta)$ and $\hat u=-\frac{\hat s}{2}(1-z)(1+y_\theta)$.
\end{widetext}
\end{document}